\documentclass[9pt,conference]{IEEEtran}
\IEEEoverridecommandlockouts

\usepackage{cite}
\usepackage{amsmath,amssymb,amsfonts}
\usepackage{algorithmic}
\usepackage{graphicx}
\usepackage{textcomp}
\usepackage{xcolor}
\usepackage{url}
\usepackage[symbol]{footmisc}
\usepackage{bm}
\usepackage{xspace}
\usepackage{lipsum}

\usepackage{graphicx}
\usepackage{subcaption}

\usepackage{booktabs}
\usepackage{siunitx}

\usepackage{pifont}
\newcommand{\cmark}{\ding{51}}
\newcommand{\xmark}{\ding{55}}

\definecolor{MyOrange}{HTML}{d66000}
\definecolor{MyGray}{gray}{0.7}

\newcommand{\ourmodelname}{\textsc{NARSiS}}
\newcommand{\ourtwostage}{NAR 2-stage}

\newcommand{\ourmodel}{\ourmodelname\xspace}
\newcommand{\ourmodelbase}{\ourmodelname\textsubscript{base}\xspace}
\newcommand{\ourmodelcodes}{\ourmodelname\textsubscript{codes}\xspace}
\newcommand{\ourmodelcont}{\ourmodelname\textsubscript{feats}\xspace}
\newcommand{\ourmodelavg}{\ourmodelname\textsubscript{avg}\xspace}

\newcommand{\mcal}[2][]{\mathcal{#2}\ifx\relax#1\relax\else^{\text{#1}}\fi}
\newcommand{\mbf}[2][]{\bm{\mathrm{#2}}\ifx\relax#1\relax\else^{\text{#1}}\fi}

\def\BibTeX{{\rm B\kern-.05em{\sc i\kern-.025em b}\kern-.08em
    T\kern-.1667em\lower.7ex\hbox{E}\kern-.125emX}}
\begin{document}

\bstctlcite{IEEEexample:BSTcontrol}

\title{Single-stage TTS with Masked Audio Token Modeling\\and Semantic Knowledge Distillation}

\author{
\IEEEauthorblockN{
Gerard I. Gállego\IEEEauthorrefmark{1}\thanks{\IEEEauthorrefmark{1}Work carried out during an internship at Dolby Laboratories.}
}
\IEEEauthorblockA{
\textit{Universitat Politècnica de Catalunya}\\
}
\and
\IEEEauthorblockN{
Roy Fejgin
}
\IEEEauthorblockA{
\textit{Dolby Laboratories}\\
}
\and
\IEEEauthorblockN{
Chunghsin Yeh
}
\IEEEauthorblockA{
\textit{Dolby Laboratories}\\
}
\and
\IEEEauthorblockN{
Xiaoyu Liu
}
\IEEEauthorblockA{
\textit{Dolby Laboratories}\\
}
\and
\IEEEauthorblockN{
Gautam Bhattacharya
}
\IEEEauthorblockA{
\textit{Dolby Laboratories}\\
}
}

 \maketitle

\renewcommand{\thefootnote}{\arabic{footnote}}

\begin{abstract}
Audio token modeling has become a powerful framework for speech synthesis, with two-stage approaches employing semantic tokens remaining prevalent. In this paper, we aim to simplify this process by introducing a semantic knowledge distillation method that enables high-quality speech generation in a single stage. Our proposed model improves speech quality, intelligibility, and speaker similarity compared to a single-stage baseline. Although two-stage systems still lead in intelligibility, our model significantly narrows the gap while delivering comparable speech quality. These findings showcase the potential of single-stage models to achieve efficient, high-quality TTS with a more compact and streamlined architecture.
\end{abstract}


\section{Introduction}
\label{sec:introduction}

In recent years, the field of speech synthesis has advanced rapidly, driven by breakthroughs in deep learning and generative models~\cite{kongHiFiGANGenerativeAdversarial2020, kimConditionalVariationalAutoencoder2021,casanovaYourTTSZeroShotMultiSpeaker2022,shenNaturalSpeechLatentDiffusion2023,juNaturalSpeechZeroShotSpeech2024}. Among various approaches, \textit{audio token modeling} has emerged as a prominent paradigm~\cite{wangNeuralCodecLanguage2023, borsosSoundStormEfficientParallel2023,kharitonovSpeakReadPrompt2023}, driven by continuous progress in \textit{neural audio codecs}~\cite{zeghidourSoundStreamEndEndNeural2021,defossezHighFidelityNeural2023,kumarHighFidelityAudioCompression2023, yangHiFiCodecGroupresidualVector2023} and \textit{large language models} (LLMs)~\cite{openaiGPT4TechnicalReport2024,llamateamLlamaHerdModels2024}. In this approach, a neural audio codec converts audio waveforms into discrete token sequences, effectively simplifying the audio modeling task by making it analogous to language modeling. Neural audio codecs enable efficient audio compression at extremely low bitrates while preserving high-fidelity reconstruction~\cite{defossezHighFidelityNeural2023,yangHiFiCodecGroupresidualVector2023,kumarHighFidelityAudioCompression2023}. Recent approaches employ \textit{residual vector quantization} (RVQ)~\cite{zeghidourSoundStreamEndEndNeural2021}, a hierarchical method that uses multiple layers of quantizers. However, audio codecs mainly emphasize low-level spectral details while lacking high-level linguistic structures, such as semantics, that can be crucial for intelligibility in the context of speech synthesis.

To address this limitation, incorporating semantic knowledge into TTS systems has proven highly effective. Many systems leverage pre-trained speech encoders~\cite{hsuHuBERTSelfSupervisedSpeech2021,chungW2vBERTCombiningContrastive2021,chenWavLMLargeScaleSelfSupervised2022} to extract semantic representations, which are then used to condition the model. These systems typically follow a two-stage process: first, generating semantic tokens from text and then producing audio tokens, each requiring separate models for training and inference. This method allows the model to generate high-level content before focusing on finer acoustic details~\cite{borsosAudioLMLanguageModeling2023}, though it introduces significant computational complexity due to the need for two decodings. Alternatively, approaches such as \textit{semantic codecs} aim to integrate semantic information directly into the audio codec~\cite{zhangSpeechTokenizerUnifiedSpeech2023, yeCodecDoesMatter2024, juNaturalSpeechZeroShotSpeech2024}. However, this often requires retraining the typically GAN-based codec, which can lead to training instability. Other methods, such as~\cite{detaiRALLE2024}, explicitly learns the attention between the audio tokens and the phonemes, which increases the complexity of data processing and model inference. 

Turning to another crucial aspect, audio token modeling in TTS can follow autoregressive (AR)~\cite{borsosAudioLMLanguageModeling2023,kharitonovSpeakReadPrompt2023} or non-autoregressive (NAR)~\cite{wangMaskGCTZeroShotTextSpeech2024} approaches. AR models generate tokens sequentially, while NAR models, through Masked Token Modeling~\cite{changMaskGITMaskedGenerative2022}, predict all tokens in parallel and iteratively refine low-confidence outputs. NAR models are significantly more efficient at inference, decoupling generation steps from sequence length and producing high-quality audio in just a few dozen steps~\cite{wangMaskGCTZeroShotTextSpeech2024}. Hybrid models~\cite{wangNeuralCodecLanguage2023} that combine AR for initial layers with NAR for subsequent layers aim to balance the trade-off between quality and efficiency.

In this work, we seek to advance the audio token modeling paradigm for speech synthesis by combining the efficiency of NAR and single-stage approaches. We maintain that incorporating semantic representations is key to enhancing TTS quality, particularly in terms of intelligibility. However, instead of introducing semantic information as an additional inference stage, which introduces inefficiencies, we propose integrating it during training through a method we term \textit{semantic knowledge distillation} (SKD). This approach distills high-level knowledge from a self-supervised speech encoder, such as HuBERT~\cite{hsuHuBERTSelfSupervisedSpeech2021}, into the model. By doing so, we also avoid relying on semantic codecs~\cite{zhangSpeechTokenizerUnifiedSpeech2023,liuSemantiCodecUltraLow2024,yeCodecDoesMatter2024}, eliminating dependencies on specific tokenizers while maintaining compatibility with state-of-the-art general audio codecs.

Our proposed speech synthesis model, \ourmodel (\textbf{NAR} \textbf{Si}ngle \textbf{S}tage TTS), is trained with an objective function that models audio and semantic tokens in parallel using a unified network. The architecture jointly learns representations at the semantic and acoustic levels, allowing semantic content to effectively guide audio token generation. The main contributions of this work are: (1) a model that significantly outperforms the single-stage baseline, especially in intelligibility, and narrows the performance gap between single- and two-stage NAR TTS models; and (2) the introduction of SKD, a novel technique for incorporating semantic knowledge during training without adding extra complexity during inference.
While our approach does not yet fully match the performance of two-stage systems, it represents a step forward in closing this gap while retaining the efficiency and simplicity of a single-stage architecture.

\section{NAR Single-Stage TTS}
\label{sec:method}

\begin{figure}[t]
    \centering
    \begin{subfigure}[t]{0.28\textwidth}
        \centering
        \includegraphics[height=7.8cm]{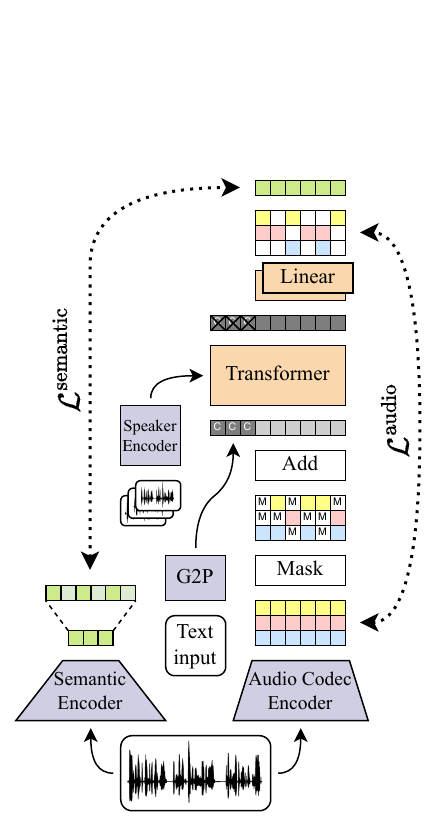}
        \caption{Training}
        \label{fig:figure1}
    \end{subfigure}
    \begin{subfigure}[t]{0.20\textwidth}
        \centering
        \includegraphics[height=7.8cm]{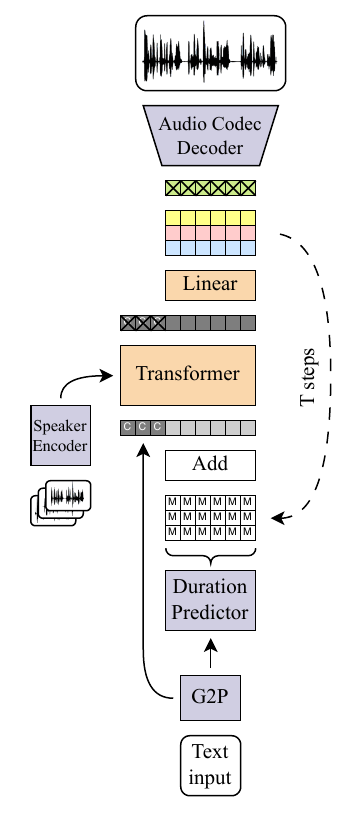}
        \caption{Inference}
        \label{fig:figure2}
    \end{subfigure}
    \caption{System overview.}
    \label{fig:bothfigures}
    \vspace{-2mm}
\end{figure}

The proposed system, named \ourmodel, is a Non-Autoregressive (NAR) Text-to-Speech (TTS) model that combines semantic and acoustic modeling in parallel, maintaining a single-stage design for enhanced efficiency. An overview of the system's architecture is provided in Fig.~\ref{fig:bothfigures}, and the key components are detailed in Sec.\ref{sec:method_arch}. Additionally, we adopt a \textit{masked audio token modeling} (MATM) approach based on MaskGIT~\cite{changMaskGITMaskedGenerative2022}, further explained in Sec.~\ref{sec:method_matm}.

\subsection{Architecture}
\label{sec:method_arch}

The core of our system is a Transformer~\cite{vaswaniAttentionAllYou2017}, which is responsible for modeling audio tokens. Additional components include an audio codec, which functions as both an audio tokenizer and detokenizer, a grapheme-to-phoneme (G2P) model that converts input text into phonemes, and a speaker encoder, which is essential for generating speech in the voice of the target speaker. Additionally, a semantic encoder is employed during the training phase, as detailed in Sec.~\ref{sec:method_skd}.

\noindent\textbf{Audio Codec}\; Speech is transformed into discrete tokens through the use of a neural audio codec, resulting in a sequence of much shorter length. The MATM is performed on these discrete audio codes. At inference time, the codec's decoder reconstructs the output waveform from the generated tokens. The audio codec consists of multiple layers of residual vector quantization (RVQ), with each layer depicted in Fig.~\ref{fig:bothfigures} using different colors.

\noindent\textbf{G2P}\; In our system, we convert the input text into phonemes, as phonemes provide a closer representation of speech sounds compared to graphemes. To perform this transformation, we employ a pre-trained Grapheme-to-Phoneme (G2P) model.

\noindent\textbf{Transformer}\; The system has a Transformer architecture inspired by MaskGIT~\cite{changMaskGITMaskedGenerative2022} as its core generative component. The input sequence of phonemes is first transformed into phoneme embeddings via a learnable embedding layer. Simultaneously, the discrete audio tokens generated by the audio codec are mapped into audio embeddings through $K$ learnable embedding layers, each corresponding to one of the layers of the RVQ. Following the approach proposed by~\cite{changMaskGITMaskedGenerative2022}, a masking mechanism is applied to the audio embeddings, where selected tokens are replaced by a learnable mask embedding. The masked embeddings are then aggregated by summing the $K$ different RVQ layers. The resulting aggregated masked audio embeddings are concatenated with the phoneme embeddings to form the input to the Transformer. This combined sequence is processed through multiple self-attention layers, where the phoneme embeddings serve to condition the model on the content. Finally, the output is projected through $K$ fully-connected layers. These projections after softmax produce the predicted audio token probabilities for the $K$ RVQ layers.

\noindent\textbf{Speaker Encoder}\; To enable multi-speaker compatibility, our system conditions the audio token generation on target speaker embeddings. These are derived through an enrollment stage, which involves selecting multiple utterances from the same speaker. A speaker embedding is extracted from each utterance using a pre-trained model. These embeddings are then averaged to create a final speaker embedding that captures the characteristics of the target speaker. This final embedding is used to condition the Transformer via Adaptive Layer Normalization (AdaLN)~\cite{xuUnderstandingImprovingLayer2019}.

\noindent\textbf{Duration Predictor}\; A key challenge in NAR systems is determining the appropriate length of the audio to be generated. To address this, we independently train a duration prediction module that estimates the required audio length based on the input text. This module operates on phonemes derived from the same G2P model and follows an approach inspired by~\cite{cambaraMapacheMaskedParallel2024}. The input phoneme sequence is passed through a learnable embedding layer to produce phoneme embeddings. A special classification token is prepended to the embeddings, similarly to~\cite{devlinBERTPretrainingDeep2019a}, and they are processed through several Transformer layers. The output corresponding to the classification token is projected through a fully-connected layer to predict the duration of the utterance in log seconds. The duration predictor is trained by minimizing a Huber loss with respect to the actual utterance duration. During inference, the length of the input sequence for the Transformer is determined based on the predicted duration.

\noindent\textbf{Semantic Encoder}\; The system includes a frozen speech encoder to extract the semantic information required for the SKD (Sec.~\ref{sec:method_skd}). This pre-trained model is used exclusively during the training phase.

\subsection{Masked Audio Token Modeling}
\label{sec:method_matm}
Our system generates audio tokens by predicting masked tokens in parallel, as opposed to the sequential approach in AR models. During training, the model learns to reconstruct masked audio tokens, with the proportion of masked tokens following a cosine schedule. The model is optimized using cross-entropy loss, computed only on the masked positions. At inference, the system begins with a fully masked token sequence and iteratively unmasks portions over multiple steps, following a similar cosine schedule. At each step, the most confident predictions among the previously masked tokens are unmasked, while the rest remain masked. This process continues until the entire token sequence is generated, which is then converted into a waveform.

\begin{table*}[t]
\caption{Objective evaluation results on LibriSpeech test-clean. VALL-E SSS uses 3\,sec enrollment based on the reported result.}
\label{tab:objective}
\centering
\begin{tabular}{l c S[table-format=3.0] S[table-format=2.1] S[table-format=1.2] S[table-format=1.2] S[table-format=1.2] c c}
\toprule
{} & {Semantic KD} & {Params. (M)} & {WER ($\downarrow$)} & {MCD ($\downarrow$)} & {SSS ($\uparrow$)} & {UTMOS ($\uparrow$)} & {Training Data} & {Hours} \\
\midrule
Ground truth & {-} & {-} & 2.2 & 0.00 &  0.80 & 4.08 & {-} & {-} \\ 
DAC recon. & {-} & {-} & 2.4 & 1.87 & 0.77 & 3.91 & {-} & {-} \\ 
\midrule
\ourmodelbase & {\xmark} & 249 & 14.6 & 6.79 & 0.55 & 3.91 & &  \\ 
\ourmodelcodes & {discrete} & 249 & 9.8 & 6.70 & 0.62 & 3.79 & & \\ 
\ourmodelcont & {continuous} & 249 & 5.9 & 6.56 & 0.60 & 3.99 & LibriTTS-clean & $245$ \\ 
\ourmodelavg  & {continuous} & 249 & 6.3 & 6.66 & 0.62 & 4.13 & & \\ 
\ourtwostage  & {\xmark} & 476 & 3.6 & 7.02 & 0.61 & 4.27 & & \\ 
\midrule
YourTTS~\cite{casanovaYourTTSZeroShotMultiSpeaker2022} & {-} & {-} & 7.1 & 8.32 & 0.50 & 3.62 & LibriTTS-clean+Pt+Fr & $474$ \\ 
VALL-E~\cite{wangNeuralCodecLanguage2023} & {-} & 370 & 5.9 & {-} & 0.58 & {-} & LibriLight & $60$k \\
StyleTTS 2~\cite{liStyleTTSHumanLevelTextSpeech2023} & {-} & {-} & 4.0 & 6.43 & 0.41 & 4.20 & LibriTTS-clean & $245$ \\ 
\bottomrule
\end{tabular}
\vspace{-2mm}
\end{table*}

\section{Semantic Knowledge Distillation}
\label{sec:method_skd}

Modeling audio tokens alone may lack higher-level semantic abstraction, so we incorporate an external self-supervised learning (SSL) speech encoder to provide semantically-rich representations from the target speech utterances. These features are extracted from a specific layer of the encoder and can either be continuous hidden representations or discrete quantized representations. To inject these semantic features into \ourmodel, we add a fully-connected layer on top of the Transformer that predicts the semantic representations. This process, which we term \textit{semantic knowledge distillation}, allows for effective semantic modeling without requiring a separate stage. The system is trained using a combined loss defined by:

\vspace{-5pt}
\begin{equation*}
    \mathcal{L} = \alpha \cdot \mathcal{L}^{\text{audio}} + \beta \cdot \mathcal{L}^{\text{semantic}}
\end{equation*}
\vspace{-12pt}

where $\mathcal{L}^{\text{audio}}$ is the cross-entropy loss on the audio tokens, and $\mathcal{L}^{\text{semantic}}$ corresponds to the loss calculated on the semantic features, depending on whether they are continuous or discrete. The terms $\alpha$ and $\beta$ are hyperparameters used to balance the contributions of these two losses. Since the sequence length produced by the audio tokenizer differs from that of the speech semantic encoder, nearest neighbor interpolation is applied to match the sequence lengths before computing the semantic loss.

For SKD, the semantic encoder provides multiple representations for experimentation:

\noindent\textbf{Discrete codes}; Using discrete semantic tokens is the standard approach in two-stage models~\cite{borsosAudioLMLanguageModeling2023,kharitonovSpeakReadPrompt2023}. The quantization process reduces low-level speaker and spectral variability, yielding a compact set of semantic tokens. In this case, a fully-connected layer projects the Transformer output to the number of discrete semantic tokens. After applying a softmax, the token probabilities are obtained. The semantic loss, $\mathcal{L}^{\text{semantic}}$, is computed using cross-entropy, comparing the predicted token probabilities with the reference semantic codes.

\noindent\textbf{Continuous features}; While discretization reduces variability, it may lead to a loss of important information. An alternative is to use continuous features directly, preserving the full richness of the original representation. Here, a fully-connected layer projects the Transformer output to the embedding dimension of the semantic encoder. In this case, the semantic loss, $\mathcal{L}^{\text{semantic}}$, is calculated using the cosine similarity implemented in~\cite{zhangSpeechTokenizerUnifiedSpeech2023}. This measures the cosine similarity along the time axis for each feature dimension.

It is important to note that SKD enables the training of a single model, rather than requiring separate models for text-to-semantics and semantics-to-audio. During inference, the fully-connected layer used to predict semantic representations is discarded, simplifying the architecture. This approach results in a more efficient and streamlined single-stage TTS system, significantly reducing both training and inference complexity compared to traditional two-stage methods.

\section{Experimental Setup}

\subsection{Implementation Details}
\label{sec:expsetup_impdetails}

Following the model architecture previously described, we provide the implementation details of \ourmodel.

\noindent\textbf{Audio tokenizer}\; We use the Descript Audio Codec (DAC)~\cite{kumarHighFidelityAudioCompression2023} to represent speech as audio tokens. DAC operates on input audio sampled at 44.1 kHz with a striding factor of 512, resulting in an audio token sampling rate of approximately 86 Hz. The codec has 9 RVQ layers and works with 1024-dimensional embeddings. The parameters of DAC are kept frozen throughout training.

\noindent\textbf{Phoneme embeddings}\; For the G2P conversion, we use a pre-trained SoundChoice model~\cite{ploujnikovSoundChoiceGraphemePhonemeModels2022}, with its parameters also kept frozen during training. It produces a set of 40 possible phonemes.

\noindent\textbf{Transformer}\; The main Transformer encoder in charge of the MATM comprises 16 layers, each with 16 attention heads, an embedding dimension of 1024, and a feed-forward hidden size of 4096, resulting in a total model size of 240\,M parameters. We use ReLU activations and a Pre-LayerNorm configuration. The audio token embedding layers are initialized with DAC’s codebooks and are fine-tuned during training. Sinusoidal positional encodings~\cite{vaswaniAttentionAllYou2017} are applied to the concatenated input embeddings.

\noindent\textbf{Speaker conditioning}\; We use the open-source~\cite{wenetcommunityWespeakerWespeakervoxceleb2024} WeSpeaker model~\cite{wangWespeakerResearchProduction2023}, with speaker embeddings extracted offline. We use an enrollment of 3 randomly selected utterances from the same speaker, each with a duration ranging from 0.5 to 10\,sec.

\noindent\textbf{Duration predictor}\; The model consists of a standard Transformer encoder of 6-layers, each with 16 attention heads, an embedding dimension of 256, and a feed-forward hidden size of 1024. We use ReLU activations and a Pre-LayerNorm configuration. It is trained with similar optimization hyper-parameters than the main model, but for 20\,k steps and a learning rate of $10^{-3}$.

\noindent\textbf{Semantic Encoder}\; We rely on a pre-trained~\cite{metaresearchFacebookHubertbasels9602021} HuBERT~\cite{hsuHuBERTSelfSupervisedSpeech2021} to extract speech semantics. We experiment with continuous and discrete representations at different layers. Specifically, we have selected the 9th layer and the average from all the layers. For the discrete case, we use a publicly available k-means clustering with 500 centroids.

\noindent\textbf{Training}\; We train \ourmodel for a total of 700\,k steps with an effective batch size of 64. We use the AdamW optimizer~\cite{loshchilovDecoupledWeightDecay2018}, with a learning rate of $10^{-4}$, $\beta_1 = 0.9$, $\beta_2 = 0.999$, and no weight decay. A polynomial decay scheduler is applied, with a learning rate warmup over 2\,k steps, a power of 0.9, and a final learning rate of $5 \cdot 10^{-7}$. Additionally, we employ Classifier-Free Guidance (CFG)~\cite{hoClassifierFreeDiffusionGuidance2021}, with a 10\% conditioning dropout, where text conditioning is replaced by a learnable token representing the unconditional mode.

\noindent\textbf{Sampling}\; Based on the sampling method of MaskGIT~\cite{changMaskGITMaskedGenerative2022}, we perform 20 steps of iterative sampling. Similar to~\cite{garciaVampnet2023}, we introduce a diversity term by adding Gaussian noise to the logits at each step where variance is linearly annealed from 3 to 0 during the sampling. We also apply CFG with a linearly decreasing guidance level from 3 to 0.75 across the sampling steps.

\noindent\textbf{Data}\; We train our models using the clean subsets of the LibriTTS corpus~\cite{zenLibriTTSCorpusDerived2019} (\textit{train-clean-100} and \textit{train-clean-360}), which together provide 245 hours of speech data. For validation, we use the \textit{dev-clean} subset. During training, we filter out utterances shorter than 0.5 seconds or longer than 10 seconds. 
To facilitate a  comparison with VALL-E~\cite{wangNeuralCodecLanguage2023}, a representative token-based TTS system, we follow their test data setting, using only samples between 4 and 10 seconds, resulting in a 2.2-hour subset from the LibriSpeech \textit{test-clean}~\cite{panayotovLibrispeechASRCorpus2015}.

\subsection{Experiments}
\label{sec:expsetup_exp}

We evaluate our proposed model by comparing it against two baselines and experimenting with different configurations of SKD using either discrete or continuous representations.

\noindent\textbf{Baselines}\; We establish two baseline models for comparison with our proposed approach. \ourmodelbase is a single-stage model without SKD. This model serves as a reference point to evaluate the effectiveness of incorporating semantic knowledge. We also build a two-stage system with the same base architecture. The first stage is a MaskGIT that generates discrete HuBERT tokens, which condition the second stage by summing the embeddings as in~\cite{borsosAudioLMLanguageModeling2023}. The difference in performance between these models is the reference for the potential improvements SKD can bring to \ourmodel.

\noindent\textbf{\ourmodelcodes}\; We train \ourmodel using SKD with discrete semantic tokens from the 9th layer of HuBERT, and clustered into 500 centroids. Prior research has shown that features from this are particularly useful for speech understanding tasks, such as speech recognition, due to their reduced speaker and spectral variability~\cite{chenWavLMLargeScaleSelfSupervised2022}. The choice of 500 centroids is based on empirical evidence~\cite{zhangSpeechTokenizerUnifiedSpeech2023, hsuLowResourceSelfSupervisedLearning2024}. For this configuration, we apply the loss described in Sec. \ref{sec:method_skd} for discrete codes, with the loss weights tuned to $\alpha=0.95$ and $\beta=0.05$ based on development set performance.

\noindent\textbf{\ourmodelcont}\; We also train \ourmodel using SKD with continuous representations from HuBERT. By leveraging non-quantized features from the 9th layer, thus preserving all information contained in this representation. The loss function used is the one described in Sec. \ref{sec:method_skd} for continuous features, with loss weights tuned to $\alpha=0.5$ and $\beta=0.5$ based on development set performance.

\noindent\textbf{\ourmodelavg}\; Inspired by~\cite{chenWavLMLargeScaleSelfSupervised2022}, showing that different Transformer layers in HuBERT capture features useful for different downstream tasks, we propose \ourmodelavg. This variant average continuous features across all 12 layers to capture a broader range of information. We use the same loss function and weights as in \ourmodelcont.

\begin{table}[tb]
\caption{Subjective MOS with $95\%$ confidence interval.}
\label{tab:subjective}
\centering
\begin{tabular}{l c S[table-format=1.2(2), separate-uncertainty, table-align-text-post=false]}
\toprule
 {} & {Semantic KD} & {MOS ($\uparrow$)} \\
\midrule
Ground truth & {-} & 4.05 \pm 0.13 \\
\ourmodelbase & {\xmark} & 3.05 \pm 0.17 \\
\ourmodelavg & {\cmark} & 3.55 \pm 0.15  \\
\ourtwostage & {\xmark} & 3.63 \pm 0.14 \\
\bottomrule
\end{tabular}
\vspace{-4mm}
\end{table}

\subsection{Evaluation Metrics}
\label{sec:expsetup_eval}

We evaluate our models using several metrics to assess various aspects of speech synthesis quality:

\noindent\textbf{WER}\; We use Word Error Rate to assess speech intelligibility by transcribing the generated speech with a pre-trained speech recognition model~\cite{metaresearchFacebookHubertlargels960ft2022}, which is also used in VALL-E to report their results.

\noindent\textbf{SSS}\; To evaluate how well the synthesized speech matches the target speaker, we compute the Speaker Similarity Score. Speaker embeddings are extracted using a publicly available~\cite{microsoftUniSpeechPretrainingRepresentations2024} WavLM model~\cite{chenWavLMLargeScaleSelfSupervised2022}, and cosine similarity is measured between embedding of the synthesized utterance and the average embedding of three different enrollment utterances.

\noindent\textbf{MCD}\; We use Mel Cepstral Distortion to quantify reconstruction error, further assessing speech quality. An open-source implementation~\cite{chenqi008Chenqi008Pymcd2022} is employed, with dynamic time warping enabled to align the generated and reference audio.

\noindent\textbf{UTMOS}\; We estimate the Mean Opinion Score (MOS) using an open-source MOS prediction system~\cite{saekiUTMOSUTokyoSaruLabSystem2022} to evaluate the overall quality of the synthesized speech.

\noindent\textbf{MOS}; In addition to objective evaluations, we conduct subjective listening tests to compare our models. Listeners were asked to rate speech naturalness, considering prosody, intonation, quality, and intelligibility. They were provided with the ground truth transcription to help evaluate word correctness. The test was carried out by 15 expert listeners and consisted of 15 evaluations.

\section{Results}
\label{sec:results}

For objective evaluation, we compare our models with three other models: YourTTS~\cite{casanovaYourTTSZeroShotMultiSpeaker2022}, StyleTTS 2~\cite{liStyleTTSHumanLevelTextSpeech2023} (both using official checkpoints), and VALL-E~\cite{wangNeuralCodecLanguage2023} (using reported results). The results are presented in Table \ref{tab:objective}. As an upper-bound reference, we report the transcription performance for ground truth audio and DAC-reconstructed audio (i.e., the output of encoding and decoding the ground truth audio, serving as the upper bound for our models).

All models incorporating SKD outperform the single-stage baseline across most metrics, demonstrating its effectiveness. Notably, intelligibility (measured by WER) improves significantly, from 14.6\% in the baseline to 9.8\% in \ourmodelcodes and further to 5.9\% in \ourmodelcont, highlighting the impact of SKD on speech intelligibility. Improvements in Speaker Similarity Score (SSS) are also noteworthy, with both discrete and continuous SKD models achieving 0.62 compared to 0.55 for the baseline. 

When comparing models utilizing SKD, those leveraging continuous features consistently outperform the discrete token-based variant across most metrics, especially in terms of WER. This result suggests that the complete information contained in the unquantized semantic features is beneficial for various aspects of the generated speech.  

Between the two models that use continuous features, \ourmodelcont slightly outperforms the average variant in terms of WER and MCD, while the latter achieves higher SSS and UTMOS scores. We select \ourmodelavg as our winning model due to its superior UTMOS performance. Notably, all models using SKD surpass YourTTS by a considerable margin and perform on par with VALL-E, despite VALL-E being an AR model trained on more than 200x more data and having a larger size (370\,M vs. 249\,M parameters).

Despite these promising results, our two-stage baseline still outperforms the single-stage models, particularly in terms of intelligibility, with a WER of 3.6\%. We hypothesize that the stronger inductive bias introduced by the intermediate semantic representation in two-stage systems helps keep the generation process more closely aligned with the underlying semantics. However, in other metrics, such as MCD and SSS, our single-stage models slightly outperform the two-stage system, and in UTMOS, the performance gap is small.

The subjective evaluation results, shown in Table \ref{tab:subjective}, further reinforce this conclusion. The human-rated MOS scores indicate no substantial perceived quality difference between our model with SKD and the two-stage baseline, whereas the single-stage baseline lags behind both.\footnote{Listening samples available at: \url{https://narsistts.github.io}}

We also measure the average runtime of both the one-stage and two-stage models. This evaluation is based on 100 runs on an A100 GPU. As shown in Table \ref{tab:runtime}, \ourmodel reduces runtime by up to 50\% compared to the two-stage baseline, particularly for longer utterances.

Overall, our best model with SKD (\ourmodelavg) significantly reduces the performance gap in intelligibility between our single-stage and two-stage models, while achieving comparable results to the two-stage model in terms of speech quality and speaker similarity.

\begin{table}[tb]
\caption{Inference runtime (in seconds).}
\label{tab:runtime}
\centering
\begin{tabular}{l S[table-format=1.2] S[table-format=1.2] S[table-format=1.2] S[table-format=1.2]}
\toprule
Utt. duration & {4s} & {8s} & {12s} & {16s} \\
\midrule
\ourtwostage & 1.37 & 1.37  & 1.37 & 2.53 \\
\ourmodelbase & 0.70 & 0.81 & 0.79 & 1.28 \\
\bottomrule
\end{tabular}
\vspace{-2mm}
\end{table}

\section{Conclusion}

In this paper, we introduced \ourmodel, a single-stage non-autoregressive TTS system that integrates semantic and acoustic token modeling in parallel. By incorporating semantic knowledge distillation, we leverage semantic representations during training, eliminating the need for intermediate stages at inference. Our experimental results demonstrate that \ourmodel significantly improves intelligibility and speaker similarity compared to a vanilla single-stage approach. While two-stage systems still exhibit better intelligibility, our model effectively narrows the performance gap and delivers competitive results in various aspects of the generated speech, highlighting the potential of single-stage models for efficient and high-quality TTS.

\bibliographystyle{IEEEtran}
\bibliography{main}

\vspace{12pt}

\end{document}